\documentclass[11pt]{emulateapj}
\usepackage{lscape}
\usepackage{graphics}

\input epsf

\def\be{\begin{equation}}\def\bea{\begin{eqnarray}}\def\beaa{\begin{eqnarray*}}
  \def\ee{\end{equation}}  \def\eea{\end{eqnarray}}  \def\eeaa{\end{eqnarray*}}

\slugcomment{\today: Not to appear in Nonlearned J., 45.}

\shorttitle{Galaxy Formation Model}
\shortauthors{Kim, Choi, and Park}

\begin{document}
\title{A Subhalo-Galaxy Correspondence Model of Galaxy Formation}
\author{Juhan Kim\altaffilmark{1,2}, Changbom Park\altaffilmark{1,3}, and Yun-Young Choi\altaffilmark{1,4}}

\altaffiltext{1}{Korea Institute for Advanced Study, Hoegiro 87,
Dongdaemun-Gu, Seoul, 130-722, Korea}
\altaffiltext{2}{kjhan@kias.re.kr}
\altaffiltext{3}{cbp@kias.re.kr}
\altaffiltext{4}{yychoi@kias.re.kr}


\begin{abstract}
We propose a model of allocating galaxies in cosmological N-body simulations.
We identify each subhalo with a galaxy, and assign luminosity and morphological type
assuming that the galaxy luminosity is a monotonic function of its host subhalo mass.  
The morphology assignment is made by using
two simple relations between subhalo mass and galaxy luminosity of different types.
One is using a constant ratio in luminosity of early (E/SO) 
and late (S/Irr) type galaxies at a fixed subhalo mass.
And the other assumes that galaxies of different morphological types but having an equal luminosity
have a constant ratio in their subhalo masses.
We made a series of comparisons of the properties of these simulated galaxies
with those of the SDSS galaxies. 
The resulting simulated galaxy sample is found to successfully reproduce
the observed local number density distribution except for in high density regions.
The luminosity function is studied as a function of local density. It was found that
the observed luminosity functions in different local density environments are
overall well-reproduced by the simulated galaxies. Discrepancy is found at the bright
end of the luminosity function of early types in the underdense regions and at
the faint end of both morphological types in very high density regions.
A significant fraction of the observed early type galaxies in voids seems to have
undergone a relatively recent star formation and became brighter.
The lack of faint simulated galaxies in dense regions may be due to the strong 
tidal force of the central halo which destroys less massive satellite subhalos around
in the simulation.
The mass-to-light ratio is found to depend on the local density in the way 
similar to that observed in the SDSS sample.
We have found an impressive agreement between out simulated galaxies and the SDSS galaxies
in the dependence of the central velocity dispersion on the local density and luminosity.
\end{abstract}


\section{Introduction}
The current galaxy formation paradigm can be characterized by 
``hierarchical clustering''.  This means that massive dark matter halos 
form by merging less massive halos and/or by accreting ambient matter, 
and also that a dark matter halo governs the evolution of the galaxy residing inside.
Most galaxies are believed to be hosted by the 
dark matter halos becuase the halos can provide a deep potential well 
for baryonic matter to condense and cool down. This leads to triggering
of the star formation;
gas sufficiently accumulated in the halo potential center
begins to experience many hydrodynamic processes,
such as radiative cooling, star-formation, supernova explosion, and chemical enrichments.
All of these processes play an important role in making the visible galaxies in the end.
Because galaxies as a building block of the large scale structures
consequently follows the evolution of their host halos over the cosmic history,
understanding gravitational evolution of dark halos is 
very important for the study of galaxy formation and evolution.

Over the past few decades, cosmological simulations have been proved 
useful for the study of structure formation.
Simulations have boosted up many investigations of the nonlinear structure
evolutions and their results have been extensively compared to the observations 
in various aspects of interest.
Many studies have reported successful recovery of observational
features of the galaxy
distribution like the two-point correlations of galaxies
\citep{conroy06,kravtsov04,berlind02}, topology \citep{park05a,park05b,gott06},
and the environmental dependence of spin distributions \citep{cervantes-sodi07}.

Detailed modeling of galaxy evolution and understanding of the galaxy properties
have been possible with the recent advent of the huge redshift surveys
such as Sloan Digital Sky Survey\footnote{http://www.sdss.org} (SDSS) and 
2dFGRS\footnote{http://www.aao.gov.au/2df/}.  These larger surveys provide 
a rich information on the formation and evolution of galaxy.
To meet a requirement to establish a relation
between simulated structures and observed galaxies
many techniques have been introduced. 
The semi-analytic model(SAM) of galaxy formation
\citep{cole94,kauffmann97,baugh06} 
is based on the hierarchical clustering of halos 
whose merging history trees are built by generating a set of
Gaussian random numbers.
The mass growth history of halos on various mass scales can be 
generated and traced by this method.
Numerous SAM parameters
are implemented in the merging history to reflect the hydrodynamical processes
of heating, cooling, star formation, and
aging the stellar populations.
Their parameter values are fine tunned for the best description 
of collective properties of observed galaxies.  
But as the number of observational contraints increases, the number of parameters
increases and the SAM becomes much complicated. 

Hydrosimulations employ direct formulations of the hydrodynamic processes (\citealt{weinberg06}). 
Two types of hydro simulations are now widely adopted;
the Lagrangian \citep{monaghan92,hernquist89} and Eulerian methods
(\citealt{tasker06,harten97}, and for comparative study, 
see \citealt{heitmann05,thacker00}).
They set gas particles or gas grids in the system of interest.
Then, gas dynamics are taken into account by calculating the hydrodynamic interactions 
between neighbor particles or by solving differential hydro equations between 
adjacent grids. 
This method has several weak points;
it suffers from the lack of resolutions in space and mass
as the N-body simulations. 
And in most cases, the star formation and supernova
explosions are far beyond the simulation resolution.
Some processes such as radiative transfer, star formation,
supernovae feedback, and initial stellar mass function,
are not thoroughly understood and
are difficult to parameterize in a well established way.

One of the offsprings of the SAM is 
the Halo Occupation Distribution (HOD; \citealt{zheng05,seljak00,berlind02}) 
model. It relates the Friend-of-Friend (FoF) halo mass to the number of 
contained subhalos 
(or galaxies) using a conditional probability, $P(N|M)$, where $M$ is the FoF halo mass
and $N$ is the number of subhalos inside the FoF halo.
This probability function is obtained from numerical simulations 
\citep{kravtsov04,berlind02,jing98} or 
from observations \citep{zehavi04,abazajian05,zheng07} fitting 
the model to the observed two-point correlation functions.

Another variant of the SAM is a one that adopts the one-to-one monotonic 
correspondence between the galaxy luminosity and subhalo mass
\citep{marinoni02,vale04,vale06,shankar06}.
This method is often called the ``correspondence model''. It is simpler
than other methods because it requires only two prerequisites;
the luminosity function of galaxies and mass function of subhalos.
Its key assumption that a more massive subhalo hosts a brighter galaxy,
is consistent with the hierarchical clustering picture.
As the halo mass grows, its luminosity is expected to grow in general 
because merging of halos may be followed by the merging of galaxies.
By matching these two functions, the subhalo mass is mapped to galaxy luminosity.

However, sometimes
a galaxy may survive the merger event because baryonic component of a galaxy 
is usually more concentrated and is more tightly bound than its dark counterpart.
These ``orphan'' galaxies \citep{gao04} which have no separate corresponding halos
could exist in the cluster regions.
Also halos of mass below a certain characteristic scale
are unable to provide baryonic matter with gravitational attraction
strong enough to resist against supernova explosions which can tear up 
the small-mass system. Even though such cases are many, 
it is sound to expect that the galaxy census is 
closely related to the population of subhalos because most of the observed galaxies 
are field galaxies who have their own dark halos.
It is worth investigating the hypothesis of the subhalo-to-galaxy correspondence 
model of galaxy formation.

In this paper, we apply the subhalo-galaxy correspondence model to a large
N-body simulation and compare the statistical properties of simulated galaxies 
with those of galaxies observed by the SDSS.
We organize this paper as follows.
In section 2 we describe our simulation and the subhalo finding method.
In section 3, we implement the subhalo-galaxy correspondence
model and show how to assign morphological types to mock galaxies.  
In section 4, the local density is introduced to quantify local environments
and the local density distributions of mock and SDSS galaxies are investigated.
We also compare the luminosity distribution of
simulated galaxies with those of the SDSS galaxies in section 5.
The environmental dependence of central velocity distributions are studied
for the mock and SDSS galaxies in section 6.
And discussions and conclusions follow in section 7.

\section{Simulation and Halo Finding}
\label{simfinding}
We have carried out a cosmological N-body simulation of the universe with 
the WMAP 3-year cosmological parameters. We ran $1024^3$ cold dark matter particles
using an improved version of the GOTPM code \citep{dubinski04}.
The code adopts a dynamical domain decomposition for the Particle-Mesh part
with a variable width of the z-directional domain slabs.
It also uses more compact and efficient oct-sibling tree walks 
reducing the computational cost for the short-range force update
which consumes about 90 \% of total run time.
As a results, this new version outperforms the previous version by about factor three 
in speed.
The simulation was run on a beowulf-type system installed 
at Korea Institute for Advanced Study. 
The linux cluster consists of 256 AMD cores and 1 tera-byte main memory. 

The simulation and cosmological parameters adopted in this study are lists in 
Table \ref{sim}.
The number of time steps in the simulations are empirically pre-determined 
to satisfy the requirement that
the maximum displacement of particles in a step should be less than the 
force resolutions which is $0.1$ times the mean interparticle separation ($d_{\rm mean}$).
The starting epochs of simulations are chosen to constrain that
any particle should not overshoot neighbors 
when the Zel'dovich displacement is made.

Subhalos are identified by the PSB method \citep{kim06}. The method applies
the FoF algorithm to identify dark matter particle groups
adopting the standard linking length, $l_{\rm link} = 0.2 d_{\rm mean}$.
Then it divides each FoF halo into subhalos.
This two-stage halo finding is in common with other methods \citep{springel01,shaw06}.
In the second stage we build a particle density field by a coordinate-free method
as usually adopted in the smoothed particle hydrodynamics \citep{monaghan92}.
This adaptive kernel implementation is intended to resolve tight clusterings of particles.
Then, we search for 26 nearest neighbors at each particle.
To construct isodensity contour we move along the neighbor positions
in a similar way used in the origingal PSB method (for details see \citealt{kim06}).
Self-bound and tidally stable subhalos are identified by measuring the tidal radius
of the subhalos and total energies of their member particles;
member candidates are selected if their distances to the center of a subhalo
are less than tidal radius of the subhalo.
Among these particles, gravitationally unbound particles are discarded. 
The resulting subhalos are used to allocate galaxies.
We call the most massive subhalo a central halo
and other subhalos satellite halos in the FoF group.
Also, the group of particles found by the FoF method are named the FoF group or FoF halo.

\begin{deluxetable*}{c|ccccccccccccc}
\tablecaption{Simulation parameters}
\tablewidth{0pt}
\tablehead{
\colhead{name}
&\colhead{$N_p$}
&\colhead{$N_m$}
&\colhead{$L_{box}$\tablenotemark{a}}
&\colhead{$N_{step}$}
&\colhead{$z_{i}$}
&\colhead{$h$}
&\colhead{$\Omega_m$}
&\colhead{$\Omega_b$}
&\colhead{$\Omega_\Lambda$}
&\colhead{$n_s$\tablenotemark{b}}
&\colhead{$b$}
&\colhead{$m_p$\tablenotemark{c}}
&\colhead{$\epsilon$\tablenotemark{d}}
}
\startdata
H2 & $1024^3$ &$1024^3$&256 & 3800 &95 & 0.732 & 0.238 &0.042&0.762&0.958&1.314&$1.0\times 10^{9}$ &$25h^{-1}{\rm kpc}$
\enddata
\tablenotetext{a}{in $h^{-1}{\rm Mpc}$}
\tablenotetext{b}{spectral power index}
\tablenotetext{c}{particle mass in $h^{-1}{\rm M_\odot}$}
\tablenotetext{d}{force resolution}
\label{sim}
\end{deluxetable*}

\section{The Subhalo-Galaxy Correspondece Model}
\label{model}
We use the monotonic one-to-one correspondence model between galaxies and subhalos;
there is one and only one galaxy in each subhalo and a more massive subhalo hosts
a more luminous galaxy. We apply this model to assign galaxies within our simulation box.

The one-to-one correspondence model is formalized in the following way.
We first measure the mass function $\Phi (M_h)$ of PSB subhalos. 
We then take the observed luminosity functions of the early (E/S0) and late (S/Irr) type
galaxies. The relation between the halo mass and the corresponding absolute magnitude
limits is given by the following integral equation.
\begin{eqnarray}
\nonumber
\int_{M_h}^\infty \Phi(M^\prime) dM^\prime &=& 
\int_{-\infty}^{\mathcal{M}_E(M_h)} \phi_E(\mathcal{M^\prime}) d\mathcal{M^\prime}\\
&+& \int_{-\infty}^{\mathcal{M}_L(M_h)} 
\phi_L(\mathcal{M^\prime}) d\mathcal{M^\prime}
,
\label{m2b}
\end{eqnarray}
where $\Phi$ is the subhalo mass function
and $\phi_E$ and $\phi_L$ are the luminosity functions of early
and late type galaxies.
In this equation we separate the full luminosity function of subhalos into 
those of early and late types to take into account the difference in mass
of the halos associated with galaxies of different morphological type.

Two models are proposed to utilize the simple mass and luminosity scaling ratios 
between different morphological types at the fixed luminosity and at the fixed mass, 
respectively. The first model uses
\begin{equation}
L_L(M_h) = L_E(\kappa M_h),
\label{fm}
\end{equation}
where $M_h$ is the subhalo mass, and
$L_L$ and $L_E$ are the luminosities of late and early types, respectively.
The equation means that at an equal luminosity the early type galaxy is $\kappa$ times
more massive than the late type galaxy.
This model is based on the findings of Park et al. (2007a) 
who have assured that the early type galaxies brighter than $M_r =-19.5$
have about $\sqrt{2}$ times higher central velocity dispersion and pairwise
peculiar velocity difference compared to the late
types of the same brightness. This implies $\kappa\approx 2$ 
if $(M_h/L)_E = \kappa(M_h/L)_L$.
Another convencing observational evidence for $\kappa=2$ can be also 
found in \citet{mandelbaum06} who obtained the same result for halos of mass greater than $10^{11}
h^{-1}{\rm M_\odot}$ from the analysis of the galaxy-galaxy weak lensing of the SDSS sample.
Equation (\ref{fm}) then makes equation (\ref{m2b}) a one-to-one relation between halo mass and 
galaxy luminosity. From now on, we name it $\kappa$ model.
The second model uses a constant factor for the luminosity rather than the mass.
This relation can be formulated as
\begin{equation}
L_L(M_h) = \beta L_E(M_h),
\label{lle}
\end{equation}
which says that late type galaxies are brighter than early type galaxies
of the same mass by a factor of $\beta$. Now we call it the $\beta$ model.

For the galaxy luminosity distribution, we use the Schechter function
\begin{eqnarray}
\nonumber
\phi(\mathcal{M}) &=& \left(0.4 \ln{10} \right)\phi^\star 10^{-0.4(\mathcal{M}-\mathcal{M^\star})
(\alpha+1)} \\
&&\exp\left[-10^{-0.4(\mathcal{M}-\mathcal{M^\star})}\right],
\end{eqnarray}
where we adopt the type-specific Schechter function parameters
$\mathcal{M}_E^\star-5\log_{10} h =-20.23$, 
$\phi_E^\star = 7.11 \times 10^{-3}h^3{\rm Mpc}^{-3} $,
and $\alpha_E=-0.53$ for early type galaxies
and 
$\mathcal{M}_L^\star-5\log_{10} h =-20.12$, 
$\phi_L^\star = 12.27 \times 10^{-3}h^3{\rm Mpc}^{-3} $,
and $\alpha_L=-0.90$ for late types in the $r$ band, which are given in Table 2 
of Choi et al. (2007) for the SDSS galaxies brighter than $\mathcal{M}_r =-19.0$
(Hereafter, we will drop the term $5 {\rm log}_{10} h$ in the absolute magnitude
and the magnitudes are all $r$--band magnitude.)
These parameter values are quite different from those given by \citet{blanton01}
and Blanton et al. (2003) mostly because we are using the type-specific functions.
Reducing the effects of the internal extinction by Choi et al. also
makes a significant difference in the $\alpha$ parameter compared to Blanton et al.'s results.  
Difference also comes from the data size (Dr4plus versus $\sim$DR1), adopted cosmology 
($\Omega_m=0.27, \Omega_\Lambda=0.73$ versus $\Omega_m=0.3, \Omega_\Lambda=0.7$), and
sample definition (volume-limited versus apparent magnitude-limited).

\begin{figure}
\plotone{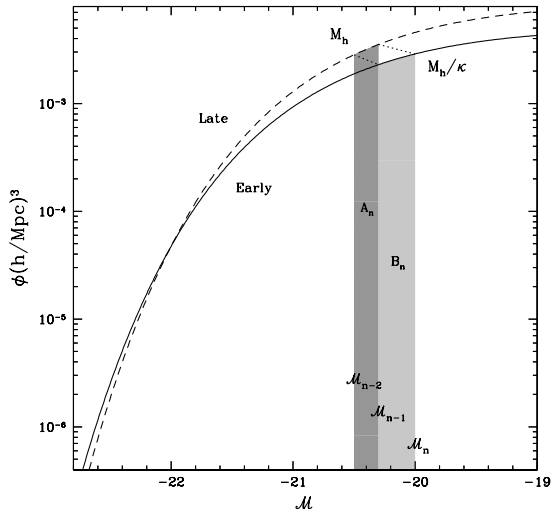}
\caption{
Illustration to derive the $M_h$--$\mathcal{M}_r$ relation for the $\kappa$.
The solid and dashed curves are the luminosity functions of early and late type galaxies,
respectively. And the dotted line connects the same mass of the two types. 
The shaded areas tagged by $A_n$ and $B_n$ are the numbers
of galaxies in the magnitude intervals between $\mathcal{M}_{n-2}$ and $\mathcal{M}_{n-1}$    
for the late type and between $\mathcal{M}_{n-1}$ and $\mathcal{M}_{n}$    
for the early type, respectively.
}
\label{kappa}
\end{figure}

The $\beta$ model can be easily solved by using the relation, 
$\mathcal{M}_L(M_h) = \mathcal{M}_E(M_h)-2.5\log_{10}{\beta}$ in equation (\ref{m2b}),
but the $\kappa$ model is a bit tricky to solve and needs a few assumptions.
The $\kappa$ model can be solved by a chain of equations whose
$n$--th term is,
\begin{eqnarray}
\nonumber
\int_{M_h/\kappa}^{M_h} \Phi(M) dM &=& \int_{\mathcal{M}_{n-2}}^{\mathcal{M}_{n-1}} 
\phi_L(\mathcal{M})d\mathcal{M}\\
&+& \int_{\mathcal{M}_{n-1}}^{\mathcal{M}_{n}} \phi_E(\mathcal{M})d\mathcal{M},
\label{eqkappa}
\end{eqnarray}
where $\mathcal{M}_{n-2} \le \mathcal{M}_{n-1}\le\mathcal{M}_{n}$ and
\begin{eqnarray}
M_h(\mathcal{M}_{L,n-2}) &=& M_h(\mathcal{M}_{E,n-1}) \equiv m\\
M_h(\mathcal{M}_{L,n-1}) &=& M_h(\mathcal{M}_{E,n}) = m/\kappa,
\end{eqnarray}
for $\kappa >1$.
Here $M_h(\mathcal{M}_L)$ and $M_h(\mathcal{M}_E)$
denote the halo masses of late type and early type galaxies of 
absolute magnitude $\mathcal{M}$, respectively.
If $M_h$, $\mathcal{M}_{n-2}$, and $\mathcal{M}_{n-1}$ are given,
then $\mathcal{M}_{n}$ can be derived.
Figure \ref{kappa} depicts the relations among the three magnitudes described above.
The solid and dashed curves show the luminosity functions of early 
and late types, respectively.
And the dotted line connects the same mass of the two morphological types.
The shaded areas denoted by $A_n$ and $B_n$ are the integrated number densities
of galaxies of magnitude between $\mathcal{M}_{n-2}$ and $\mathcal{M}_{n-1}$ 
for the late type and between $\mathcal{M}_{n-1}$ and $\mathcal{M}_{n}$ 
for the early type, respectively.
Therefore, the sum of the two shaded areas should be equal to the integrals of 
the subhalo mass function (the left-hand side of Eq. \ref{eqkappa}).
By making a stride to the next mass scale which is smaller by $1/\kappa$ times,
we can get a chain of equations and derive the scaling
relation between the mass and the luminosity.
To solve this chain of equations
we set initial conditions under the plausible assumption
that the late-type contributions to the number density at the high-mass (or 
bright) end
are negligible  compared to those of early types (or $0\simeq A_1,A_2 \ll B_1,B_2$).
This setting is quite fair from the fact that early type galaxies dominate 
the bright end population of galaxies.
Then, we are able to solve the series of equation (\ref{eqkappa}) 
from the initial conditions where
only the early-type contribution to number density is dominant.

\begin{figure}
\plotone{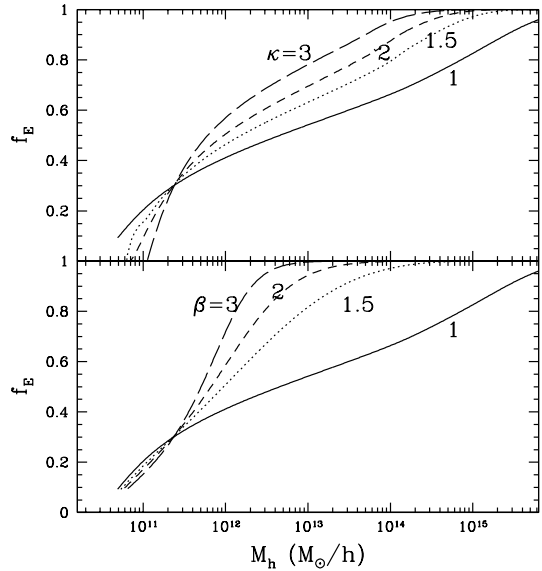}
\caption{
Early type fractions as a function of host subhalo mass.
Each curve shows the fractional distributions of early type galaxies
for $\beta=1$,
1.5, 2, and 3 ({\it lower panel}) and for $\kappa=1$,
1.5, 2, and 3 ({\it upper panel}) models. The model parameter values are attached to the curves.
}
\label{fe}
\end{figure}

Now we investigate the early type fractions as a function of the host subhalo mass.
The probability of a galaxy of mass $M_h$ to be early type is given by
\begin{equation}
f_E(M_h) = {\phi_E (\mathcal{M}_E|M_h)\over \phi_E{(\mathcal{M}_E|M_h)}
+\phi_L{(\mathcal{M}_L|M_h)}},
\end{equation}
where the denominator is the luminosity function of all galaxies of mass $M_h$
and the numerator is the luminosity function of early types of mass $M_h$.
Figure \ref{fe} shows the results for the $\kappa$ and $\beta$ models.
Using $f_E$  we are able to randomly assign morphological type to each mock galaxy
in accordance with its halo mass.
For a larger value of $\beta$,
the early type galaxy tends to dominate the population down to lower mass scales ($M_c$)
and below $M_c$ $f_E$ drops more rapidly.
But for the $\kappa$ model, those changes in $f_E$ are not so much steep.
Also, in this plot, we note that
a more dramatic change of the $f_E$ in the $\beta$ model
than the $\kappa$ model.

\begin{figure}
\plotone{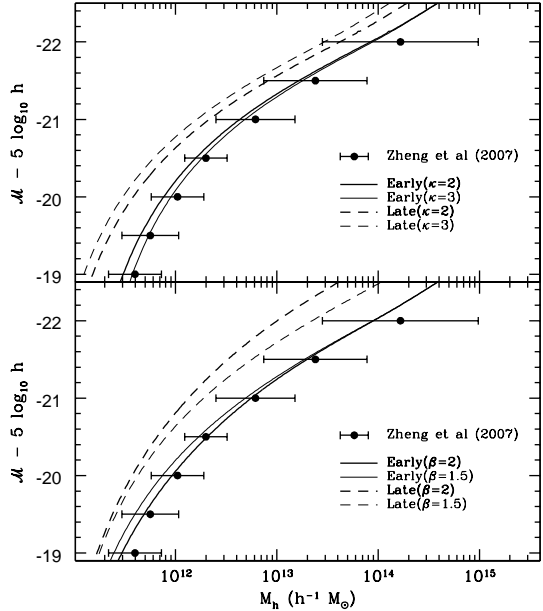}
\caption
{
Scaling relation between halo mass and galaxy magnitude in the $\kappa$ ({\it upper panel}) 
and $\beta$ ({\it lower panel}) models.
In the lower panel, {\it solid} lines shows the dependence of the modeled early type magnitude on
the host subhalo mass for $\beta=2$ ({\it thick}) and for $\beta=1.5$ ({\it thin}).
And in the upper panel, we show the mass-to-magnitude relations for $\kappa=2$ ({\it thick})
and $\kappa=3$ ({\it thin}) models.
{\it Dashed} curves show the relations of late type galaxies.
Filled circles are the best-fitting results
of the SDSS observations to the HOD model in \citet{zheng07}.
And the errorbar is the width of the cut-off profile of the step-like function
in the HOD model (for details, please see their paper).
}
\label{m2m}
\end{figure}

Figure \ref{m2m} shows the relation between the subhalo mass $M_h$ and absolute magnitude 
$\mathcal{M}$ of early ({\it solid lines}) and late ({\it dashed lines}) type galaxies derived 
from equation (\ref{m2b})
and the luminosity functions of the SDSS galaxies
in the $\kappa$ ({\it upper}) and $\beta$ ({\it lower panels}) models. 
Only shown are the cases of $\beta=2$ (thick lines) and 1.5 (thin lines) and
$\kappa=2$ (thick lines) and 3 (thin lines).
For a comparison, 
the characteristic minimum mass of the central subhalos at each absolute magnitude 
limit estimated by Zheng et al. (2007) based on the HOD model is given as filled circles.
They obtained the magnitude-to-mass relation of 
the central subhalo but ignored the morphological types of galaxies.
The HOD model is quite consistent with our $M_h$--$\mathcal{M}$ relation for the
early type galaxies  with $\beta= 2$, and $\kappa= 2$ and 3.

\begin{figure}
\plotone{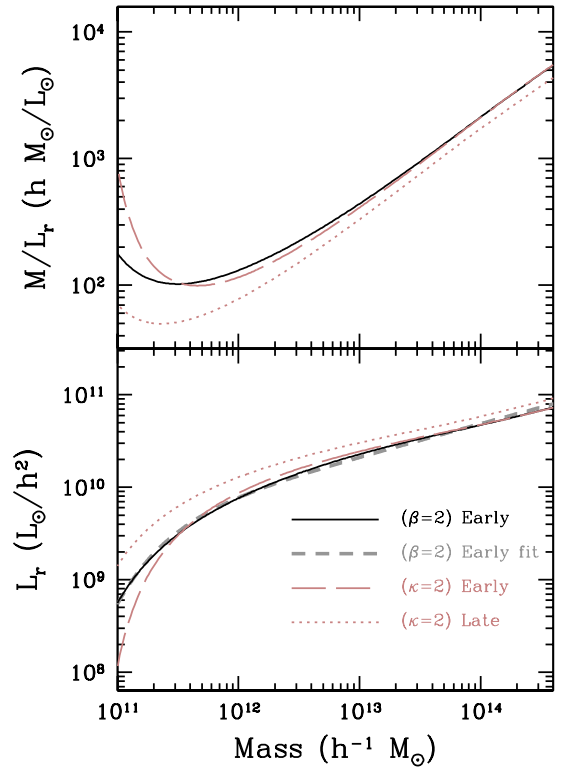}
\caption
{
The scaling relation between the halo mass and the galaxy luminosity 
({\it lower panel}) or mass-to-light ratio ({\it upper panel}) in the $\beta=2$ 
and $\kappa=2$ model.
In the lower panel, the {\it solid} line shows the derived luminosity of
early type mock galaxies as a function of the subhalo mass and the {\it short dashed}
line is the best-fitting curve for it.
Also the early ({\it long dashed}) and late ({\it dotted}) luminosity distributions
are shown for the $\kappa=2$ model.
In the upper panel, the mass-to-light ratio of the mock galaxies is shown
by the line of the same type as shown in the lower panel.
But we do not plot the fitting result in upper panel.
}
\label{ml}
\end{figure}

In the lower panel of Figure \ref{ml} shows the mass versus luminosity relation (solid line)
for the early type galaxies. Here we adopt $\mathcal{M}_{\odot}=4.64-5\log_{10} h$ \citep{blanton07}
to transform the magnitude to luminosity in the $r$--band filter.
It can be noted that galaxy luminosity drops steeply at the low mass end and 
rises in a power law at the high mass end.
Taking into account these features, we propose a function 
\begin{equation}
L({M}) = {{M}^\star\over \Psi_{ml}} \left({{M}^\star\over {M}}\right)^{\gamma-1}
e^{-\left({{M}^\star\over M\phantom{\star}}\right)},
\label{analml}
\end{equation}
as a fitting function for the halo mass versus early-type galaxy luminosity relation.
We apply the $\chi$--square fitting to the relation and 
obtain the best-fit parameter values of $\Psi_{ml} =38.3 h M_\odot /L_\odot$, 
$M^\star=2.04\times 10^{11}h^{-1}{\rm M_\odot}$, and $\gamma=0.644$ for the early type
in the $\beta=2$ model.
In the $\kappa=2$ model, $\Psi_{ml} =39.8(20.3) h M_\odot /L_\odot$,
$M^\star=4.83(1.52)\times 10^{11}h^{-1}{\rm M_\odot}$,
and $\gamma=0.667(0.719)$ for early (late) type.
Only the fitting result for the early type in the $\beta=2$ model is shown by a short-dashed line.
The top panel shows the mass-to-light ratios of the early (solid curve) types
in the $\beta=2$ model. The long-dashed and dotted curves show those ratios of early 
and late type galaxies in the $\kappa=2$ model.
In the $\beta=2$ model,
there is a upturn of the mass-to-light ratio of early types
around $M_u=3\times 10^{11}h^{-1}{\rm M_\odot}$
(for those measured  in other bands, see \citealt{bosch03,yang03,eke06}),
which means the star formation in galaxies is strongest at the mass scale.
A fitting formula for the mass-to-light ratio is derived from equation (\ref{analupsilon}),
\begin{equation}
\Upsilon \equiv M/L = \Psi_{ml} \left({M\over M^\star}\right)^{\gamma}
e^{\left({M^\star\over M\phantom{\star}}\right)}.
\label{analupsilon}
\end{equation}
Using this equation,
the mass scale corresponding to the minimum mass-to-light ratio can be related
to the shape parameters as $M_{u}=M^\star/\gamma$ and the minimum mass-to-light ratio is 
$\Upsilon_{u}(E) = \Psi_{ml} ({e/\gamma})^{\gamma}\simeq 100$ for early type galaxies
and $\Upsilon_{u}(L) \simeq 50$ for late types in the $\beta=2$ model.

According to our result the mass-to-light ratio of the brightest galaxies reaches much
higher than $10^3$. 
But this value is too high compared to those reported in literature.
For example, \citet{mandelbaum06} noted that a mass-to-light of early types reaches $M_h/L_r=674$
for their brightest samples ($-22.5 \le \mathcal{M} \le -22$ in Tab. 4 of their paper)
of the SDSS lensing galaxies.
The discrepancy may be explained by the fact that they are typically the brightest cluster galaxies. 
In our PSB halo identification method all mass not assigned to satellite galaxies is
assigned to the central subhalo after the boundness check,
which makes the central subhalo as massive as the whole 
cluster and their mass-to-light ratio very large
({\it cf.} Fig. 10 in \citealt{vale06} and Fig. 4 in \citealt{tinker05}). 
Then, the group or cluster mass-to-light ratio is the one to be compared
at the very high mass scales.

\section{Environmental Effects on the Galaxy Distribution}
\label{environ}
\subsection{Definition of the Local Density}
For a comparative study of the environmental effects on the spatial and luminosity
distributions of observed and simulated galaxies,
a quantitative measure of the local environment is needed.
To minimize the parameterizations and maximize the spatial resolution 
we use the spline kernel to obtain the smooth galaxy number density field. 
It is given by
\begin{displaymath}
W(q) = \left\{ \begin{array}{ll}
        \left(1-{3\over2} q^2 +{3\over4} q^3\right)/(\pi h_s^3) &\textrm{for }0<q\le1,\\
        \left(2-q\right)^3/(4\pi h_s^3) &\textrm{for } 1<q\le2,\\
        0 &\textrm{otherwise},
		\end{array}\right.
\end{displaymath}
where $q\equiv r/h_s$.
The kernel is centrally weighted more than the Gaussian, 
and has only one parameter, $h_s$.
It is smooth to the second order and has a finite tail out to $2h_s$.
Throughout this paper we set $h_s=d_{20}/2$, where $d_{20}$ is the distance to the
twentieth nearest neighbor galaxy.
Due to these features and to the adaptive nature of $h_s$
the resulting galaxy density map better represent the observed distribution of galaxies
than the Gaussian, particularly in voids and clustered regions.
To compare the observed and mock galaxy samples,
it is necessary to make the number density of the density tracers the same.
We use galaxies brighter than $\mathcal{M} = -20$ as the galaxy density tracers.
This selection differs slightly from that adopted by \citet{park07a}
who chose $-21<\mathcal{M}<-20$. We removed the bright cut to better resolve the centers of
clusters where galaxies brighter than $L_\star$ are concentrated.

\subsection{Number Density Distributions}
The local density at galaxy positions is measured in our SDSS volume-limited
samples.  The density estimate is corrected for the boundary effects when the kernel sphere 
of radius $d_{20}$ overlaps with the survey boundary.
The late type galaxies having axis ratio $b/a <0.6$ are removed from the sample
to reduce the effects of internal absorption on luminosity.
After this exclusion, we give the weight of 1/0.505 to late type galaxies
to correct the galaxy number density for the missing inclined ones, where
$0.505$ is the fraction of late types  with $b/a>0.6$ in a sample
containing galaxies down to $\mathcal{M}=-17.5$ (the CM sample of Choi et al. 2007).

The correction for boundary effect is unnecessary for mock galaxies
since their distribution is periodic in all directions.
The redshift distortions effects caused by the peculiar velocities are mimicked
by shifting the mock galaxies along the $x$ axis using $x^\prime=x + v_x/H_0$, 
where $v_x$ is the $x$-component of peculiar velocity and $H_0$
is the hubble constant.

\begin{figure}
\plotone{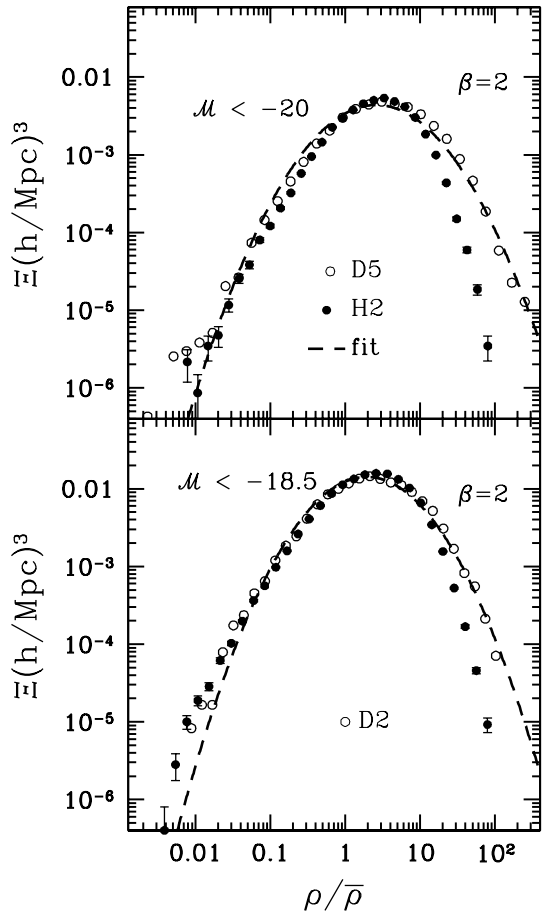}
\caption{
Distributions of local overdensities for the sample of mock ({\it filled circles})
and SDSS galaxies ({\it open circles})
in the magnitude ranges
of $\mathcal{M} <-20$ ({\it upper}) and $\mathcal{M}< -18.5$ ({\it lower} panels)
in the $\beta=2$ model.
The dashed line shows the log-normal fit to observations.
}
\label{densim}
\end{figure}

\begin{figure}
\plotone{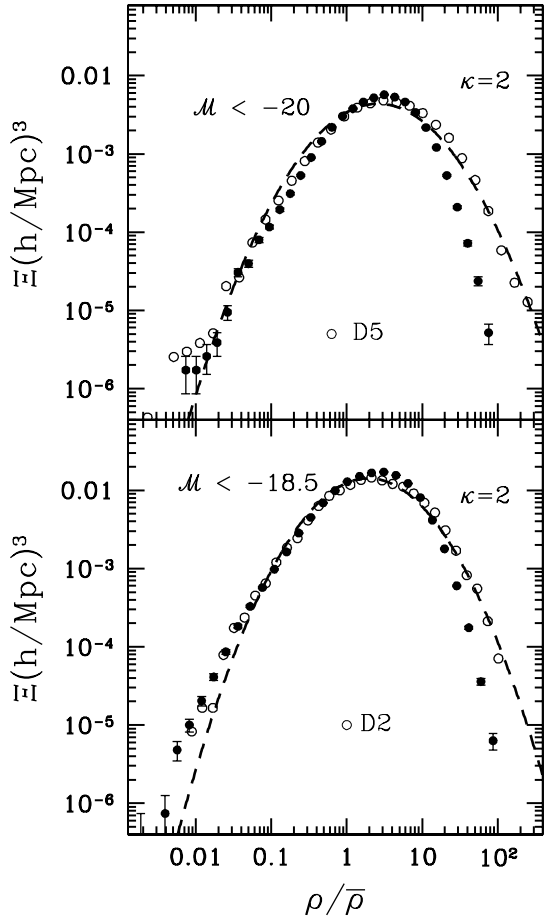}
\caption{
Similar to Fig. \ref{densim} but for the $\kappa=2$ model.
}
\label{kdensim}
\end{figure}

Figure \ref{densim} 
shows the distribution of local density ({\it open circles})
at the location of galaxies in two volume-limited SDSS samples with absolute magnitude
limits of $-20.0$ (D5) and $-18.5$ (D2) for the $\beta=2$ model.
Also plotted are the local density distributions of simulated galaxies selected by 
the same magnitude-limit criteria.
It can be noted that the local density distributions of the SDSS galaxies and
the corresponding mock galaxies match closely to each other at all densities
except for at high densities.
Also those distributions for the $\kappa=2$ model are shown in Figure \ref{kdensim} with the
same criteria.


\citet{ostriker03} employed a log normal function to fit
the one-point distribution of dark matter density and luminosity density 
in their hydro-simulation.
A spherical top-hat filter of constant radius is adopted in their study.
They found a good agreement between the simulated density distributions 
and the log normal distribution
while the distribution of luminosity density shows a poor fit.
We check how well the galaxy density distribution, instead of dark matter or luminosity,
 is described by the non-normalized log normal function
\begin{equation}
\Xi(\Delta)\equiv{dN(\Delta)\over d\log_{10}\Delta}
={A\over\sqrt{2\pi}\sigma} e^{-\left(\ln \Delta-\mu\right)^2/2\sigma^2},
\end{equation}
where $\Delta \equiv \rho/\bar{\rho}$ and $A$ is the amplitude of the distribution.
The mean number density of galaxies is $n= A \log_{10}e$.
The best-fitting results are shown with dashed lines in each panel of Figure \ref{densim}
and the fitting parameter values for various magnitude-limit ($\mathcal{M}_{{\rm lim}}$)
samples are listed in Table \ref{locdenfit}.
As can be noted, brighter galaxies tend to be located at higher local densities.
Also the local density distribution becomes broader for brighter subsamples if
viewed in the linear scale of overdensity 
or has nearly a constant width ($\left<\sigma\right> \simeq 1.29$) 
in log-scale.

\begin{deluxetable}{lc|ccc}
\tablecaption{The log normal function parameters of the local density distributions of the SDSS
galaxies}
\tablewidth{0pt}
\tablehead{
{$\mathcal{M}_{{\rm min}}$}
&{sample}
&{$A (h^{-1}{\rm Mpc})^{-3}$}
&{$\mu$}
&{$\sigma$} }
\startdata
$-18.5$&D2 &$4.53\times10^{-2}$ &0.662&1.27\\
$-20$&D5 &$1.47\times 10^{-2}$ &0.949&1.34\\
$-20.5$&D5 &$7.44\times 10^{-3}$ &1.11&1.30\\
$-21$&D5 &$2.76\times 10^{-3}$ &1.23&1.26
\enddata
\label{locdenfit}
\end{deluxetable}

\begin{figure}
\plotone{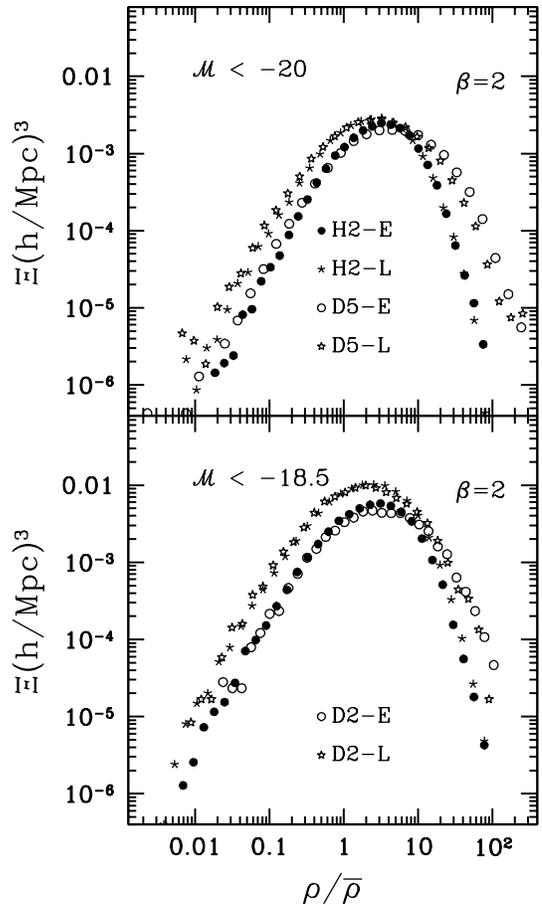}
\caption{
Local overdensity distributions for the two morphological types in the $\beta=2$ model.
The distributions of local overdensity of mock and SDSS galaxies (D5 and D2) are shown
for two magnitude-limit samples $\mathcal{M} <-20$ ({\it upper}) 
and $\mathcal{M} <-18.5$ ({\it lower panels}).
{\it Open cirlces} and {\it open stars} mark the distributions of overdensity 
of the early type and late type SDSS galaxies.
And {\it filled circles} and {\it skeletal stars} are those of mock galaxies
for the corresponding types.
We tag each panel with the magnitude criteria of the sample in the upper-left corner.
}
\label{dentype}
\end{figure}

\begin{figure}
\plotone{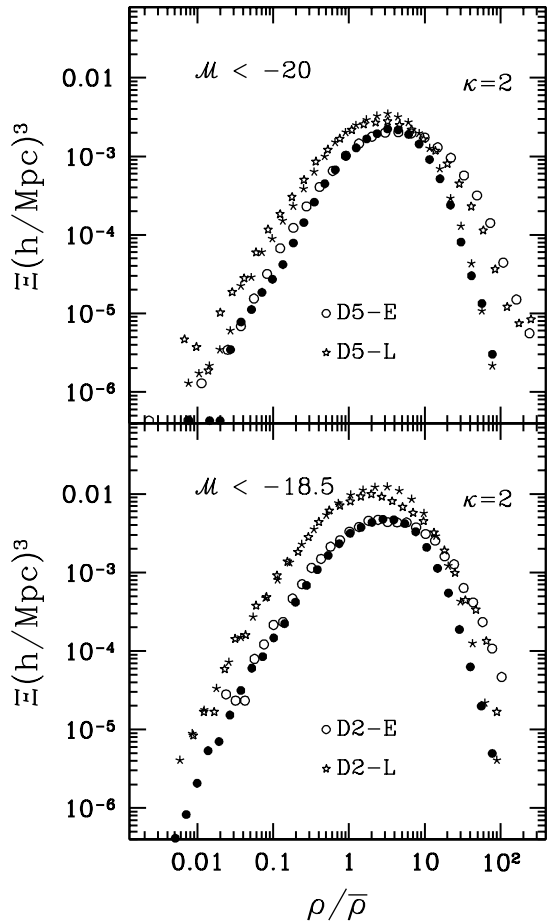}
\caption{
Same as Fig. \ref{dentype} but for the $\kappa=2$ model.
}
\label{kdentype}
\end{figure}

Figure \ref{dentype} and \ref{kdentype} shows the distribution of local density at locations 
of early (circles) and late (stars) type galaxies in the $\beta=2$ and $\kappa=2$ models, respectively. 
Open symbols are for the SDSS samples and the rest are for the simulated galaxies.
The local density distributions of the simulated early and late type galaxies are
very well matched with observations at low and intermediate densities 
($\rho/\bar{\rho}<10$) even though there is only one input parameter in our model. 
Late type galaxies dominate these regions. 
From these figures, it appears that the galaxy morphology depends 
only on the halo mass and does not directly depend on environment.
But the number density analysis 
alone is not so sufficient to draw any definite conclusion of the environmental effect.
We will further investigate this effect by comparing the luminosity functions 
of the SDSS with those of the mock galaxies in various environments in the next section.
In the high density regions our model gives too few galaxies, which is probably
due to the insufficient resolution of the simulation to maintain small subhalos
within clusters and to the lack of gas physics.


\section{Luminosity Function}
\label{lum}
There have been several works reporting the environmental dependence
of the galaxy luminosity function
(Park et al. 1994 and 2007a in observations and \citealt{mo04} in the HOD model).
It was found that the characteristic galaxy luminosity ($L_\star$)
is an increasing function of the local density,
and the faint-end slope ($\alpha$) of the luminosity function 
is insensitive to the local density.
However, Park et al. (2007a) used 
a spline kernel weighting to estimate local densities, and 
\citet{mo04} (also \citealt{cooray05}) use a spherical top-hat filter 
of a constant radius.
For a quantitative comparison between observations and models
the same density estimation scheme is required.

In the previous section, we studied the one-point distribution of the local density
at galaxy locations for each morphology sample of galaxies brighter than a certain
absolute magnitude limit. In this section, we investigate the distribution of 
the absolute magnitude of early and late type galaxies located in different local
density environment. This gives us environment and morphology-specific luminosity 
function of galaxies. In our galaxy morphology assignment scheme we only use the 
observed total luminosity function and the parameter $\beta$ (or $\kappa$), 
the ratio of luminosity (or mass) of the early and late types at a fixed mass 
(or luminosity). 
It will be interesting to see 
how accurately our model reproduces the observed luminosity functions at different
local densities and for different morphological types.
For this comparison we use the luminosity functions measured from the D3 sample of
Park et al. (2007a) because this absolute magnitude-limited sample covers both
bright and faint magnitudes well, relative to other volume-limited samples.
The fitting of the measured luminosity functions to the Schechter formula is
carried out by using the MINUIT 
\footnote{http://wwwasdoc.web.cern.ch/wwwasdoc/minuit/minmain.html}
packages which employ the maximum likelihood method.

\begin{figure}
\plotone{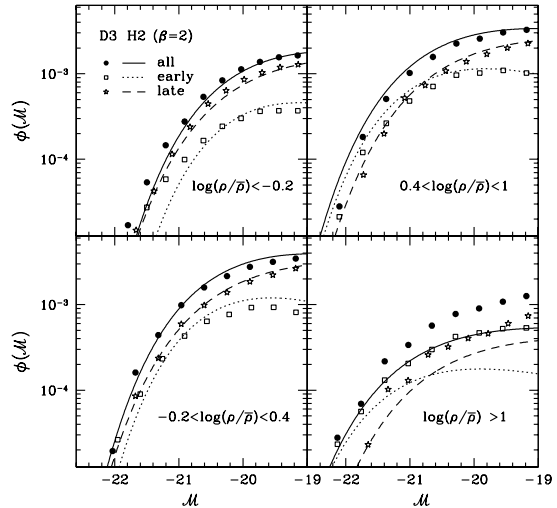}
\caption{
Luminosity functions of mock ({\it curves}) and SDSS ({\it symbols}) galaxies in the $\beta=2$ model.
Each line shows the luminosity functions of the mock galaxies;
{\it solid lines} for total, 
{\it dotted lines} for early type galaxies, and
{\it dashed lines} for late type galaxies.
For comparison,
we show the luminosity densities of the SDSS subsamples divided by the same density cuts.
The sample selection criteria are based on 
the log scale of local density and are written in the lower-right corner of each panel.
}
\label{complf}
\end{figure}

\begin{figure}
\plotone{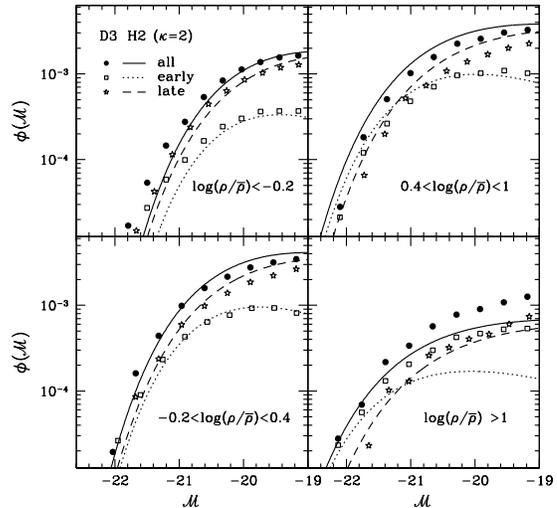}
\caption{
Similar to Fig. \ref{complf} but for the $\kappa=2$ model.
}
\label{kcomplf}
\end{figure}

Figure \ref{complf} and \ref{kcomplf} show the comparison of the luminosity functions
in four different local density regions
in the $\beta=2$ and $\kappa=2$ models.
The resulting environment and morphology-specific luminosity function 
reproduces the observations surprisingly well. However, there are notable disagreements
for two cases. At very high densities ($\rho/\bar{\rho}>10$) the abundance of faint
early type galaxies are significantly low in the simulation. This is again probably due 
to the lack of small subhalos that were destroyed in high density environment.
Another problem is seen at low densities ($\rho/\bar{\rho}<1$), where the early type
galaxies are too few in the bright end of luminosity function.
It seems that the morphology transformatin to bright early types in underdense regions 
through close interactions and mergers is more efficient in the nature than 
in our model (\citealt{park08} for observational evidence).
In the high density regions, the $\kappa=2$ model describes the number 
distribution of faint late type galaxies better than the $\beta=2$ model
while this $\kappa$ model shows a slight overestimation in the population of 
faint late types in the mean
fields ($0.4<\log(\rho/\bar{\rho})<1$) compared to the observation.

\begin{figure}
\plotone{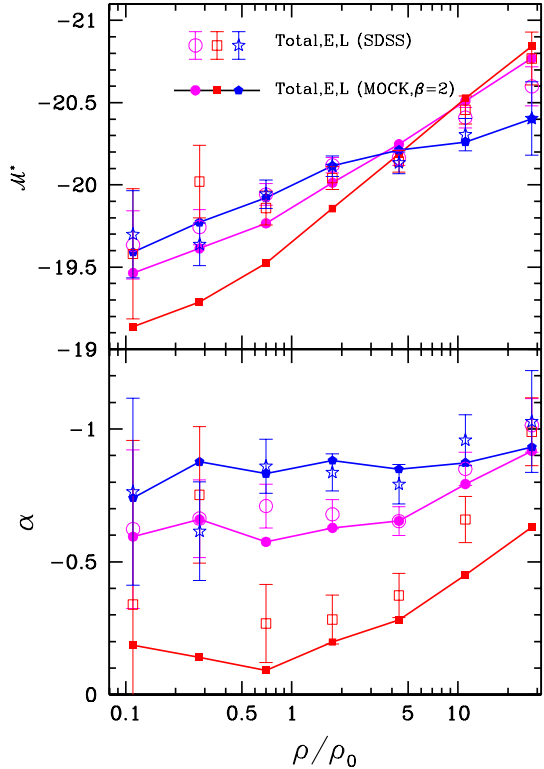}
\caption{
Dependence of the Schechter parameter values, 
$\mathcal{M}^\star$ ({\it top}) and $\alpha$
({\it bottom} panel), on the local density
for the SDSS D3 ({\it open} symbols) and Mock ({\it filled} symbols) samples
in the $\beta=2$ model.
{\it Circles} mark the fitting values of full galaxy samples and {\it boxes}
are for the early type samples.
{\it Open} stars and {\it filled} hexagons are for the late types of
the SDSS and mock samples, respectively.
In this plot we add the error bars only to the SDSS fitting values
and connect only the mock results with solid lines for clarity.
}
\label{mstaralpha}
\end{figure}

\begin{figure}
\plotone{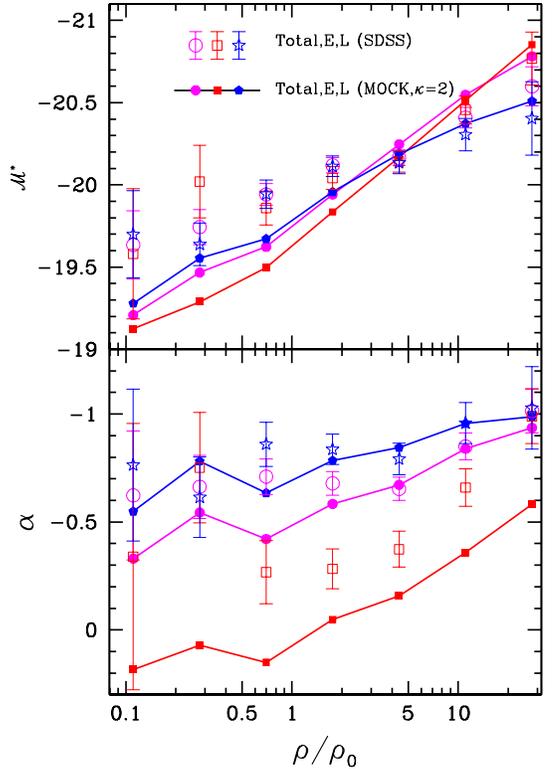}
\caption{
Similar to Fig. \ref{mstaralpha} but for $\kappa=2$ model
}
\label{kmstaralpha}
\end{figure}

Figure \ref{mstaralpha} and \ref{kmstaralpha}
compare the parameter of the Schechter function best fit 
to the SDSS data (open symbols) and simulated galaxy samples (filled symbols) as a function
of local density in the $\beta=2$ and $\kappa=2$ models. 
The dependence of both $\mathcal{M}_{r*}$ and $\alpha$ on 
local density is qualitatively well-reproduced by the simulation.
However, the characteristic magnitude $\mathcal{M}_{*}$ of the early types is 
significantly fainter at $\rho/\bar{\rho}<2$ in the simulation. This is due to paucity
of bright early type galaxies in low density regions, which is mentioned above.
The parameter $\alpha$ of the simulation is quite different from the observation
at $\rho/\bar{\rho}>2$ for early type galaxies. This is again due to the flat faint end
slope of the luminosity function of the simulated early type galaxies.


\section{Central velocity dispersion}
\begin{figure}
\plotone{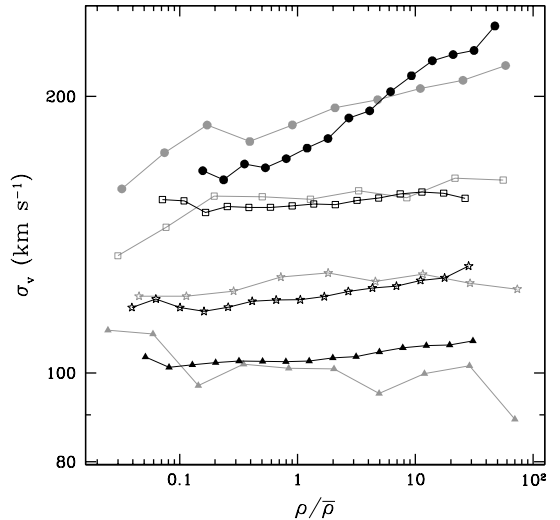}
\caption{
Median distribution of central velocity dispersions of SDSS (gray symbols) galaxies
and scaled one-dimensional velocity dispersions of Mock (black symbols) galaxies
of early types in the four magnitude-limit samples divided by four magnitude cuts.
The magnitude ranges of the subsamples are
$\mathcal{M} < -20.5$,
$-20.5<\mathcal{M} < -20.1$,
$-20.1<\mathcal{M} < -19.3$, and
$-19.3<\mathcal{M} < -19$
from top to bottom lines, respectively.
}
\label{sigv2den}
\end{figure}

Figure \ref{sigv2den} shows the central velocity dispersion
of early type galaxies in the D3 sample.
The gray symbols with connecting lines are the relations between $\sigma_v$
and the local density for early type galaxies in four subsamples
with absolute magnitude limits of 
$-19.0 > \mathcal{M}>-19.3$ (bottom curve).
$-19.3 > \mathcal{M}>-20.1$, 
$-20.1 > \mathcal{M}>-20.5$,  and
$\mathcal{M}<-20.5$ (top curve with filled circles).
The curves delineate the median value of $\sigma_v$ in each local density bin,
which monotonically increases as luminosity increases.
We also plot the scaled one-dimensional velocity dispersion of the early type mock galaxies
with black symbols.
Because the observed velocity dispersion is obtained by sampling the inner part ($\le 1''.5$)
of the galaxy,
we have to apply a scaling factor to the mock galaxy velocity dispersions.
Because it is irrelevant in this paper to derive the exact value of the scaling factor
and apply it to the related analysis,
we make a rough estimation of the relation between $\sigma_{los}$ and $V_v$,
where $\sigma_{los}$ is the line-of-sight aperture velocity dispersion and
$V_v$ is the virial velocity.
According to \citet{lokas01}, the resulting scaling ratio of the velocity dispersions,
$f_c(\equiv \sigma_{los}/V_v)$, is 
\begin{equation}
0.5\lesssim f_c \lesssim 3
\label{fc}
\end{equation}
for the acceptable ranges of the velocity anisotropy 
and the concentration parameter in $0<(r_a/R_v)<1$,
where $r_a$ is the comoving aperture radius.
Here we assume that $V_v = \sigma_v$, where $\sigma_v$ is the three-dimensional
velocity dispersion and $V_v \equiv \sqrt{2GM_v/R_v}$.
In the range of $f_c$ written in equation (\ref{fc}),
we simply set
$f_c=0.8$ ($\mathcal{M}<-20.5$), 
$f_c=1$ ($-20.5 < \mathcal{M}<-20.1$),
$f_c=1.1$ ($-20.1 < \mathcal{M}<-19.3$), and
$f_c=1.2$ ($-19.3 < \mathcal{M}<-19.0$) to match for the mean amplitudes
of velocity dispersions for each mock and SDSS sample pair.
At a fixed luminosity the velocity dispersion of the early type mock galaxies
increases as the local density increases for bright galaxies
and the slopes are decreasing for the faint sample.
It is reassuring that this trend is exactly the kind of phenomenon
found in the observation ({\it gray symbols}). 
While the switch of the slope in the faintest SDSS samples is not clearly observed 
in the mock sample (it shows a nearly flat slope on the local environments),
the change of the slope is similar to each other.
Both observation and our model show that the mass-to-light ratio of
the early type galaxies is a function of environment, and that this dependence
in turn is a function of luminosity.
The mass-to-light ratio of the early types decreases as the local density increases
for galaxies brighter than about $\mathcal{M}_*$, but increases for those
fainter than about $\mathcal{M}_*+1$.

\section{Summary \& Conclusion}
\label{summary}
We have proposed a model to assign galaxy luminosity and morphology to the dark subhalos 
directly identified in cosmological N-body simulations.
It is assumed that the galaxy luminosity is a monotonic function of its host halo mass.
In the $\kappa$ model we assume
that the halo masses of early and late type galaxies of equal luminosity have a constant ratio.
Another alternative model is the $\beta$ model which assumes the constant 
luminosity ratio between equal-mass galaxies of different types.
This model has been proposed by
\citet{marinoni02} who found that 
the observed $B$--band luminosity function of galaxies can be reproduced from the
PS function assuming double power-law mass-to-light ratios and 
derived halo occupation numbers.
It has been expanded by \citet{vale04,vale06} who adopted
the satellite halo mass functions and directly link subhalos to the observed galaxies.
In this paper, we have introduced the ratio of luminosity of the early
and late morphological type galaxies at a given halo mass, and derived
type-specific mass-to-light ratios as a function of subhalo mass.
They are used to assign luminosity and morphology to subhalos.
The mass-to-light ratio of the early type galaxies derived in this way has
a minimum value of $\Upsilon_u\simeq100$ at the scale of 
$M_u\simeq3\times10^{11}h^{-1}{\rm M_\odot}$ in the $\beta=2$ model.
The mass-to-light ratio starts to exponentionaly increase below $M_u$ 
and increases in a power law above $M_u$.

We use the large-scale background galaxy number density as an environmental parameter. 
The smooth galaxy density field is obtained by
using the adaptive spline kernel which enabled us to resolve crowded regions well.
The local density distribution of the SDSS galaxies is well described by
the log normal function.
The obtained log normal parameter values 
indicate that brighter galaxies tend to locate in dense regions.
The local density distribution of the simulated galaxies is quite similar to that of
the SDSS galaixes in voids and moderate-density regions for both morphological 
types.  
The underestimation of galaxy population in clusters
is thought to be due to the evaporation of subhalos
by the dynamical friction and tidal stripping.
This can also explain the discrepance in the luminosity function of the
early type galaxies in high density regions.

Recently, \citet{gott06} measured the genus statistic from a large sample of 
the SDSS galaxies and compared it
with those of mock galaxies created by the three distinctive methods 
such as the semi-analytic galaxy formation model
applied to the Millennium run \citep{springel05}, a hydrodynamic simulation, 
and the subhalo-galaxy correspondence model adopted in this paper.
It was found that the observed topology of large scale structure was best 
reproduced by the subhalo-galaxy correspondence model even though
other models were also consistent with the observation.
However, the observed topology was marginally inconsistent with all simulations
in the sense that it showed a strong meatball topology at the significance level
of $2.5 \sigma$ at the scale studied. The prominence of the isolated high density
regions in the observation seems to be due to the Sloan Great Wall which was
a dominant structure in the sample analyzed.

We have found an impressive agreement between our simulated galaxies and the
SDSS galaxies in the dependence of the central velocity dispersion on the local
density and luminosity. The early type galaxies tend to have higher $\sigma_v$
or higher mass in high density regions at a given luminosity when they are
brighter than about $\mathcal{M}_*$. In other words, these bright galaxies
tend to become relatively fainter in high density regions at a given halo mass. 
This interesting dependence of the mass-to-light ratio on environment
was successfully reproduced by our subhalo-galaxy correspondence model
of galaxy formation. A more detailed study of this phenomenon will be
presented in a forthcoming paper.

\acknowledgments
CBP acknowledges the support of the Korea Science and Engineering
Foundation (KOSEF) through the Astrophysical Research Center for the
Structure and Evolution of the Cosmos (ARCSEC).
Funding for the SDSS and SDSS-II has been provided by the Alfred P. Sloan
Foundation, the Participating Institutions, the National Science
Foundation, the U.S. Department of Energy, the National Aeronautics and
Space Administration, the Japanese Monbukagakusho, the Max Planck 
Society, and the Higher Education Funding Council for England.
The SDSS Web Site is http://www.sdss.org/.

The SDSS is managed by the Astrophysical Research Consortium for the
Participating Institutions. The Participating Institutions are the
American Museum of Natural History, Astrophysical Institute Potsdam,
University of Basel, Cambridge University, Case Western Reserve University,
University of Chicago, Drexel University, Fermilab, the Institute for
Advanced Study, the Japan Participation Group, Johns Hopkins University,
the Joint Institute for Nuclear Astrophysics, the Kavli Institute for
Particle Astrophysics and Cosmology, the Korean Scientist Group, the
Chinese Academy of Sciences (LAMOST), Los Alamos National Laboratory,
the Max-Planck-Institute for Astronomy (MPIA), the Max-Planck-Institute
for Astrophysics (MPA), New Mexico State University, Ohio State University,
University of Pittsburgh, University of Portsmouth, Princeton University,
the United States Naval Observatory, and the University of Washington.

The authors would like to acknowledge the use of 
the linux cluster, QUEST, at the Korea Institute for Advanced Study (KIAS).
Its huge computing power is indispensible for the study, and we thank 
the system managers for their efforts in providing a stable and comfortable 
computation resources during the simulation and subsequent analysis.


\begin{thebibliography}{}
\itemindent -5mm
\bibitem[Abazajian et al.(2005)]{abazajian05}
	Abazajian, K., Zheng, Z., Zehavi, I., Weinberg, D.H., Frieman, J.A.,
	Berlind, A.A., Blanton, M.R., Bahcall, N.A., Brinkmann, J., 
	Schneider, D.P., \& Tegmark, M. 2005, ApJ, 625, 613
\bibitem[Baugh(2006)]{baugh06}
	Baugh, C.M. 2006, Report on Progress in Physics, 69, 3101
\bibitem[Berlind \& Weinberg(2002)]{berlind02}
    Berlind, A.A. \& Weinberg, D.H. 2002, ApJ, 575, 587

\bibitem[Blanton et al.(2001)]{blanton01}
	Blanton et al. 2001, AJ, 121, 2358
\bibitem[Blanton \& Roweis(2006)]{blanton07}
	Blanton, M.R. \& Roweis, S. 2007, ApJ, 133, 754
\bibitem[van den Bosch et al.(2003)]{bosch03}
	van den Bosch, F.C., Yang, X., \& Mo, H.J. 2003, MNRAS, 340, 771
\bibitem[Cervantes-sodi et al.(2007)]{cervantes-sodi07}
	Cervantes-Sodi, B., Hernandez, X., Park, C., \& Kim, J. 2007, arXiv/0712.0843
\bibitem[Cole et al.(1994)]{cole94}
    Cole, S., Aragon-Salamanca, A., Frenk, C.S., Navarro, \& Zepf, S.E. 1994,
	MNRAS, 271, 781
\bibitem[Conroy et al.(2006)]{conroy06}
	Conroy, C.C., Wechsler, R.H., \& Kravtsov, A.V. 2006, ApJ, 647, 201
\bibitem[Cooray(2005)]{cooray05}
	Cooray, A. 2005, MNRAS, 363, 337
\bibitem[Dubinski et al.(2004)]{dubinski04}
	Dubinski, J., Kim, J., Park, C., \& Humble, R. 2003, New Astronomy, 9, 111
\bibitem[Eke et al.(2006)]{eke06}
	Eke, V.R., Baugh, C.M., Cole, S., Frenk, C.S., \& Navarro, J.F. 2006, MNRAS, 370, 1147
\bibitem[Evans \& Wilkinson(2000)]{evans00}
	Evans, N.W. \& Wilkinson, M.I. 2000, MNRAS, 316, 929
\bibitem[Gao et al.(2004)]{gao04}
	Gao, L., De Lucia, G., White, S.D.M., \& Jenkins, A. 2004, MNRAS, 352, L1
\bibitem[Gott et al.(2006)]{gott06}
	Gott, J.R., Hambrick, D.C., Vogeley, M.S., Kim, J., Park, C., Choi, Y.-Y., Cen, R., Ostriker, J.P., 
	\& Nagamine, K. 2006, ApJ, in press
\bibitem[Harten, A.(1997)]{harten97}
	Harten, A. 1997, Journal of Computational Physics, 135, 260
\bibitem[Hernandez et al.(2007)]{hernandez07}
	Hernandez, X., Park, C., Cervantes-Sodi, B., \& Choi, Y.-Y. 2007, MNRAS, 375, 163
\bibitem[Hernquist \& Katz(1989)]{hernquist89}
	Hernquist, L. \& Katz, N. 1989, ApJS, 70, 419
\bibitem[Jing et al.(1998)]{jing98}
	Jing, Y.P., Mo, H.J., \& B\"orner, G. 1998, ApJ, 494, 1
\bibitem[Kauffmann et al.(1997)]{kauffmann97}
	Kauffmann, G., Nusser, A., \& Steinmetz, M. 1997, MNRAS, 286, 795
\bibitem[Kim \& Park(2006)]{kim06}
	Kim, J. \& Park, C. 2006, ApJ, 639, 600
\bibitem[Kravtsov et al.(2004)]{kravtsov04}
	Kravtsov, A.V., Berlind, A.A., Wechsler, R.H., Klypin, A.A., 
	Gottl\"ober, S., Allgood, B., \& Primack, J.R. 2004, ApJ, 609, 35
\bibitem[Lokas \& Mamon(2001)]{lokas01}
	Lokas, E.L. \& Mamon, G.A. 2001, MNRAS, 321, 155
\bibitem[Mandelbaum et al.(2006)]{mandelbaum06}
	Madelbaum, R., Seljak, U., Kauffmann, G., Hirata, C.M., \& Brinkmann, J. 2006, 
	MNRAS, 368, 715
\bibitem[Marinoni \& Hudson(2002)]{marinoni02}
	Marinoni, C. \& Hudson, M. 2002, ApJ, 569, 101
\bibitem[Mo et al.(2004)]{mo04}
	Mo, H.J., Yang, X., van den Bosch, F.C., \& Jing, Y.P. 2004,
	MNRAS, 349, 205
\bibitem[Monaghan(1992)]{monaghan92}
	Monaghan, J.J. 1992, Annual Review of Astronomy and Astrophysics, 30, 543
\bibitem[Heitmann et al.(2005)]{heitmann05}
    Heitmann, K., Ricker, P.M., Warren, M.S., \& Habib, S. 2006
	ApJS, 160, 28
\bibitem[Ostriker et al.(2003)]{ostriker03}
	Ostriker, J.P., Nagamine, K., Cen, R., \& Fukugita, M. 2003, ApJ, 597, 1
\bibitem[Park et al.(2005a)]{park05a}
	Park, C., Kim, J., \& Gott, J.R. 2005, ApJ, 633, 1
\bibitem[Park et al.(2005b)]{park05b}
	Park, C., Choi, Y.-Y., Vogeley, M.S., Gott, J.R., Kim, J., Hikage, C.,
	Matsubara, T., Park, M.-G., Suto, Y., \& Weinberg, D.H. 2005, 633, 11
\bibitem[Park et al.(2007a)]{park07a}
	Park, C., Choi, Y.-Y., Vogeley, M., Gott, J.R., \& Blanton, M.R.  2007, ApJ, 658, 898
\bibitem[Park et al.(2008)]{park08}
	Park, C., Gott, J.R., \& Choi, Y.-Y. 2008, ApJ, in press
\bibitem[Sakamoto et al.(2003)]{sakamoto03}
	Sakamoto, T., Chiba, M., \& Beers, T.C. 2003, A\&A, 397, 899
\bibitem[Seljak (2000)]{seljak00}
    Seljak, U. 2000, MNRAS, 318, 203
\bibitem[Shankar et al.(2006)]{shankar06}
	Shankar, F., Lapi, A., Salucci, P., Zotti, G.De, \& Danese, L. 2006, ApJ,
	643, 14
\bibitem[Shaw et al.(2006)]{shaw06}
	Shaw, L.D., Weller, J., Ostriker, J.P., \& Bode, P. 2006, ApJ, 646, 815
\bibitem[Sheth \& Tormen(1999)]{sheth99}
	Sheth, R.K. \& Tormen, G. 1999, MNRAS, 308, 119
\bibitem[Sheth et al.(2001)]{sheth01}
	Sheth, R.K., Mo, H.J., \& Tormen, G. 2001, MNRAS, 323, 1
\bibitem[Springel et al.(2001)]{springel01}
	Springel, V., White, S.D.M., Tormen, G., \& Kauffmann, G. 2001, MNRAS, 328, 2001
\bibitem[Springel et al.(2005)]{springel05}
	Springel, V. et al. 2005, Nature, 435, 629
\bibitem[Tasker \& Bryan(2006)]{tasker06}
	Tasker, E.J. \& Bryan, G.L. 2006, ApJ, 641, 878
\bibitem[Thacker et al.(2000)]{thacker00}
	Thacker, R.J., Tittley, E.R., Pearce, F.R., \& Couchman, H.M.P. 2000, MNRAS,
	319, 619
\bibitem[Tinker et al.(2005)]{tinker05}
	Tinker, J.L., Weinberg, D.H., Zheng, Z., \& Zehavi, I. 2005, ApJ, 631, 41
\bibitem[Vale \& Ostriker(2004)]{vale04}
	Vale, A. \& Ostriker, J.P. 2004, MNRAS, 353, 189
\bibitem[Vale \& Ostriker(2006)]{vale06}
	Vale, A. \& Ostriker, J.P. 2006, MNRAS, 371, 1173
\bibitem[Vale \& Ostriker(2007)]{vale07}
	Vale, A. \& Ostriker, J.P. 2007, astro-ph/0701096
\bibitem[Weinberg et al.(2006)]{weinberg06}
    Weinberg, D.H., Colombi, S., Dave, R., \& katz, N. 2006, astro-ph/0604393
\bibitem[Weller et al.(2005)]{weller05}
	Weller, J., Ostriker, J.P., Bode, P., \& Shaw, L. 2005, MNRAS, 364, 823
\bibitem[Zehavi et al.(2004)]{zehavi04}
	Zehavi, I., et al. for the SDSS Collaboration 2004, ApJ, 608, 16
\bibitem[Zentner et al.(2005)]{zentner05}
	Zentner, A.R., Berlind, A.A., Bullock, J.S., Kravtsov, A.V., \& Wechler, R.H. 2005,
	ApJ, 624, 505
\bibitem[Zheng et al.(2005)]{zheng05}
    Zheng, Z., Berlind, A.A., Weinberg, D.H., Benson, A.J., 
	Baugh, C.M., Cole, S.C.,
	Dave, R., Frenk, C.S., Katz, N., \& Lacey, C.G. 2005, ApJ, 633, 809
\bibitem[Zheng et al.(2007)]{zheng07}
	Zheng, Z., Coil, A.L., \& Zehavi, I. 2007, astro-ph/0703457
\bibitem[Yang et al.(2003)]{yang03}
	Yang, X., Mo, H.J., \& van den Bosch, F.C. 2003, MNRAS, 339, 1080


\end{thebibliography}
\end{document}